\documentclass[prb,twocolumn,aps,fixfloats]{revtex4}
\usepackage{graphicx}
\usepackage{bm}
\usepackage{amsmath,amssymb,mathtools}
\usepackage{hyperref}
\usepackage{subfigure}
\usepackage{float}
\usepackage{latexsym}
\usepackage{soul}
\usepackage{color}
\usepackage{xcolor}
\usepackage{enumerate}
\usepackage{pdfpages}
\usepackage{tikz}
\usepackage{graphicx}
\usepackage{bm}
\usepackage{amssymb} 
\usepackage{amsmath}
\usepackage{subfigure}
\usepackage{relsize}

\begin{document}
\newcommand{\s}{\scriptscriptstyle}
\newcommand{\ba}{\mkern-6mu}
\newcommand{\uu}{\uparrow \uparrow}
\newcommand{\ud}{\uparrow \downarrow}
\newcommand{\du}{\downarrow \uparrow}
\newcommand{\dd}{\downarrow \downarrow}
\newcommand{\ket}[1] { \left|{#1}\right> }
\newcommand{\bra}[1] { \left<{#1}\right| }
\newcommand{\bracket}[2] {\left< \left. {#1} \right| {#2} \right>}
\newcommand{\vc}[1] {\ensuremath {\bm {#1}}}
\newcommand{\tr}{\text{Tr}}
\newcommand{\Trans}{\ensuremath \Upsilon}
\newcommand{\Refl}{\ensuremath \mathcal{R}}

\title{Scattering of electron from a disk in 2D electron gas: full cross section, transport cross section,
and the interaction correction}

\author{Nathan L. Foulk}
\author{M. E. Raikh}
\affiliation{Department of Physics and
Astronomy, University of Utah, Salt Lake City, UT 84112}

\begin{abstract}
It is known that the presence of the Fermi sea modifies the scattering of an electron from a point-like impurity. This is due to the Friedel oscillations of the electron density around the impurity. These oscillations create an additional scattering potential for incident electrons. The closer the energy of the incident electron to the Fermi level, the stronger the
additional scattering. We study this effect for the case when the impurity is not point-like but rather a hard disk, with a radius much bigger than the de Broglie wavelength. We start with a careful examination of the full and transport cross sections from an extended target. Both cross sections approach their limiting values upon increasing the wave vector of the incident electron. We establish that the transport cross section saturates much faster than the full cross section. With regard to the interaction correction, we establish that it vanishes for the full cross section, while for the transport cross section, it is enhanced compared to the case of a point-like scatterer.
\end{abstract}

\maketitle

\section{Introduction}
The static polarization operator, $\Pi({\bf q})$, of the 2D electron gas contains a singular correction,\cite{stern1967}
$\frac{M}{\pi\hbar^2}\left(q-2k_{\s F}\right)^{1/2}$, near $q=2k_{\s F}$. Here $M$ is the electron mass and $k_{\s F}$ is the Fermi momentum. This singular behavior (Kohn anomaly) translates into  the interaction
corrections to the thermodynamic characteristics of the 2D gas,\cite{Chubukov} such as effective mass. These corrections
exhibit singular behavior as a function of temperature, $T$.
Transport characteristics of the 2D gas, such as conductivity, also acquire singular interaction corrections
in the ballistic regime, $T\tau \gg 1$, where $\tau$ is the scattering time. In this regime, multiple scattering of electrons
by the impurities can be neglected, while modification of the potential of individual impurities due to the Kohn anomaly\cite{Gold1986} yields a correction to the scattering cross section proportional to $T$.
This mechanism of the anomalous temperature dependence was
pointed out in Ref. \onlinecite{Gold1986}. It was subsequently elaborated upon
in Refs. \onlinecite{Rudin1997,Zala,Gornyi2006}. Consideration of Refs.~ \onlinecite{Rudin1997,Zala,Gornyi2006}
 led to the following lucid prescription
for incorporating electron-electron interactions  into the calculation of transport.

An impurity potential, $U_{\text{imp}}({\bf r})$, causes a perturbation of the electron density
\begin{equation}
\label{Friedel}
n({\bf r})-n_0=-\nu_0 \frac{\sin(2k_{\s F}r)}{2\pi r^2}\int d{\bf r}U_{\text{imp}}({\bf r}),
\end{equation}
around it. Here $\nu_0=\frac{M}{\pi \hbar^2}$ is the density of states. The Friedel oscillations  Eq.~(\ref{Friedel})
translate into an additional scattering potential for incident electrons. The correction to the scattering amplitude due to this potential depends dramatically on the energy, $\varepsilon$, of the incident electron measured from the Fermi level, $E_{\s F}$.
Perturbative calculation of this correction\cite{Zala} indicates that it is maximal
within the angular interval $\sim \Big|\frac{\varepsilon}{E_{\s F}}-1\Big|^{1/2}$
near the backscattering condition. The relative magnitude of this correction is also
$\sim \Big|\frac{\varepsilon}{E_{\s F}}-1\Big|^{1/2}$. Thus, for a typical value $|\varepsilon-E_{\s F}|\sim T$,
the relative interaction correction to the net scattering cross section can be estimated as $\frac{T}{E_{\s F}}$.

The calculations in Refs. \onlinecite{Rudin1997,Zala,Gornyi2006} were carried out for point-like scatterers. Namely, it was assumed that their size is much smaller than the de Broglie wave length, $\frac{2\pi}{k_{\s F}}$.
In the present paper, we extend the theory\cite{Rudin1997,Zala,Gornyi2006} to the scatterers of the arbitrary
size, $a$. For this purpose, we first analyze the Friedel oscillations from the disk and also study the behavior
of the full and transport cross sections from the disk in the absence of interactions. Then we incorporate
interactions and study corresponding corrections to these cross sections.

\section{Friedel oscillations from the wall}
\begin{figure}
  \includegraphics[scale=0.25]{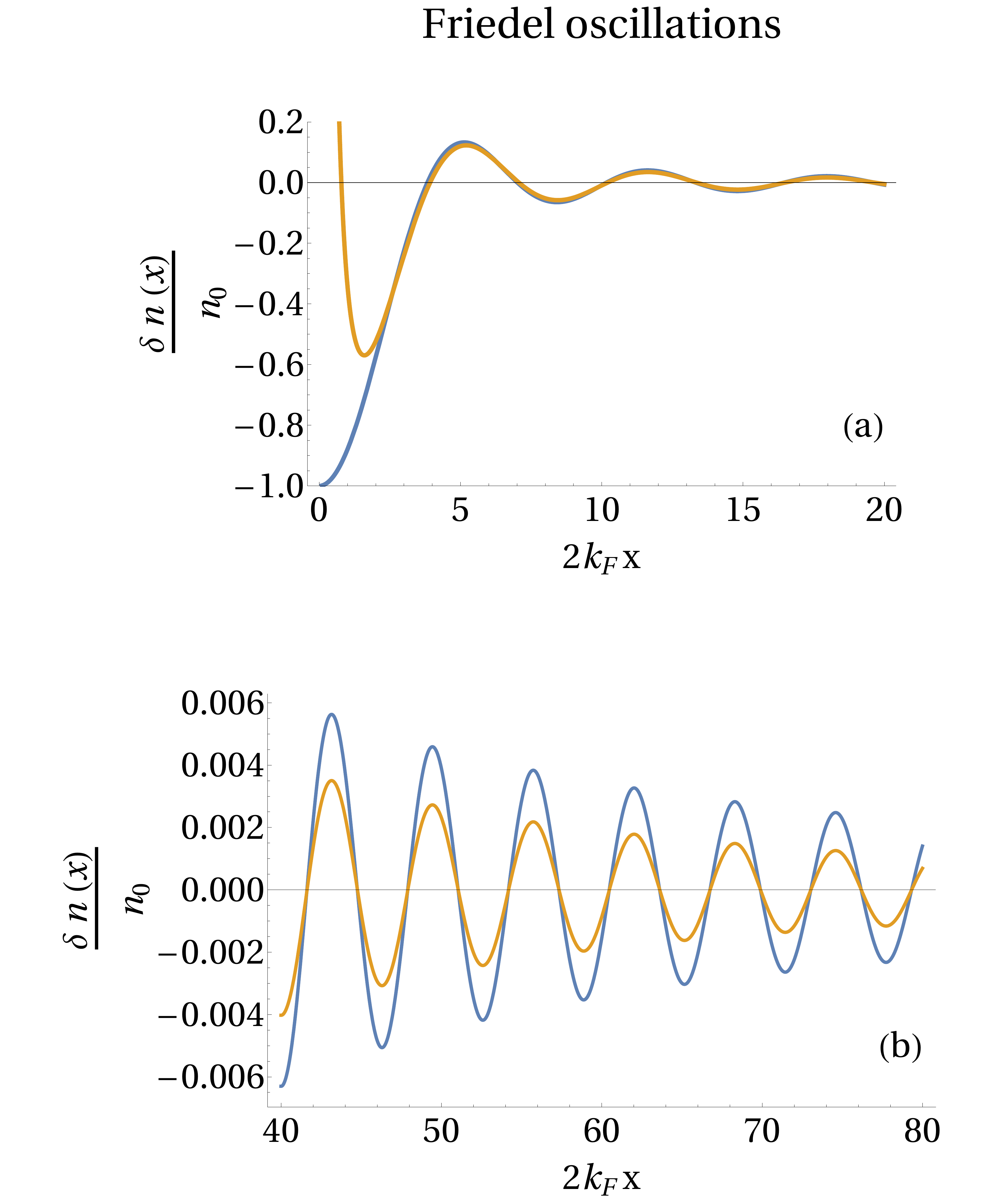}
  \caption{(Color online) (a) blue curve: Friedel oscillations from the wall are plotted from Eq.~(\ref{Wall});
  yellow curve: Friedel oscillations from the disk are plotted from Eq.~(\ref{deltaN}) for $k_{F}a = 3$.
  (b) same dependencies as in (a) are plotted for large values of $2k_{\s F}x$. Oscillations from the disk
  fall of faster than the oscillations from the wall.
  \label{figFriedel}}
\end{figure}

In terms of formation of the Friedel oscillations, at small distances, $(r-a)\ll a$,
the scatterer can be viewed as a hard wall. Then the electron density does not
depend on $a$. To calculate this density it is sufficient to substitute into
the definition
\begin{equation}
\label{density}
n({\bf r})=\sum_{\bf k}\Theta\Bigl(E_{\s F}-\frac{\hbar^2{\bf k}^2}{2M}\Bigr)\vert\Psi_{\bf k}({\bf r})\vert^2,
\end{equation}
the wave functions, $\Psi_{\bf k}=e^{ik_yy}\sin k_xx$, which turn to zero at the wall.
Here $\Theta(z)$ is a step-function.
Performing the integration over the components of the wave vector, we obtain
\begin{equation}
\label{Wall}
n(x)-n_0=-n_0\frac{J_1(2k_{\s F}x)}{k_{\s F}x},
\end{equation}
where $J_1(z)$ is the first-order Bessel function. These oscillations are shown in Fig.~\ref{figFriedel}. The concentration is zero at $x=0$.
At large distances, $k_{\s F}x \gg 1$, the relative correction to the density is small and falls off as $\left( k_{\s F}x \right)^{-3/2}$, which is intermediate between $\left( k_{\s F}x \right)^{-2}$ in 2D and $\left( k_{\s F}x \right)^{-1}$ in 1D.

\subsection{Friedel oscillations from a hard disk}
To calculate the radial dependence of the electron density,
\begin{equation}
\label{density}
n(r)\propto \sum_{k,m}\Theta\left(E_{\s F}- \frac{\hbar^2k^2}{2M} \right)
\left[ R_{m,k}^{(0)}(r)\right]^2,
\end{equation}
we use the semiclassical form of the eigenfunctions
\begin{equation}
\label{semiclassical}
R_{m,k}^{(0)}(r)=\frac{\sin\Bigl[\int\limits_a^r dr'\left(k^2-\frac{m^2-1/4}{r'^2}\right)^{1/2}\Bigr]}
{\left(k^2r^2-m^2+\frac{1}{4}  \right)^{1/4}}.
\end{equation}
The oscillating part of $n(r)$ is determined by the states with energies
close to the Fermi level, i.e. with $k$ close to $k_{\s F}$. This allows
to expand the argument of sine in $R_{m,k}^{(0)}$ with respect to $\delta k =k-k_{\s F}$
as follows

\begin{multline}
\label{EXPANSION}
\left(k^2-\frac{m^2-\frac{1}{4}}{r'^2}\right)^{1/2}
\!\!\\
=\left(k_{\s F}^2-\frac{m^2-\frac{1}{4}}{r'^2}\right)^{1/2}\!\!-\frac{k_{\s F}\delta k}{\left(k_{\s F}^2-\frac{m^2-\frac{1}{4}}{r'^2}\right)^{1/2}     }.
\end{multline}
It is also sufficient to set $k=k_{\s F}$ in the prefactor of $R_{m,k}^{(0)}$. Then the
integration over $\delta k$ yields

\begin{align}
\sum_{m=-\infty}^{\infty}&\Biggl\{1+\Biggl(\frac{k_{\s F}^2a^2-m^2}{k_{\s F}^2r^2-m^2}\Biggr)^{1/2} \Biggr\}\nonumber\\
\label{sum1}
&\times\frac{ \sin \Bigl[2\int\limits_a^r dr'\left( k_{\s F}^2-\frac{m^2}{r'^2}   \right)^{1/2}\Bigr]}
{k_{\s F}^2\left(r^2-a^2\right)}.
\end{align}

The expansion Eq.~(\ref{EXPANSION}) is justified for $\delta k \ll  k_{\s F}$.
On the other hand, characteristic $\delta k$ in the integration is $\delta k \sim \left(r-a   \right)^{-1}$. Thus, the result Eq.~(\ref{sum1})  applies for distances $(r-a)\gg k_{\s F}^{-1}$. This result describes the oscillating part of $n(r)$ within a numerical factor.

As we will see later, the main contribution to the sum Eq.~(\ref{sum1}) comes from
large momenta, $m\gg 1$, but still with $m \ll  ka$. In this domain, one can expand the
integrand in Eq.~(\ref{sum1}) and perform the integration over $dr'$, which yields

\begin{equation}
\label{expansion}
\int\limits_a^r dr' \left(k_{\s F}^2-\frac{m^2}{r'^2}   \right)^{1/2}
\approx k_{\s F}(r-a)-\frac{m^2}{2k_{\s F}}\left(\frac{1}{a}-\frac{1}{r}   \right).
\end{equation}
We see that the condition $m \ll k_{\s F}a$ ensures that the second term in Eq.~(\ref{expansion}) is much smaller than the first term.
On the other hand, the second term in Eq.~(\ref{expansion}) defines characteristic
$m \sim \left( \frac{k_{\s F}ar}{r-a}    \right)^{1/2}$. This value is smaller than
$k_{\s F}a$ under the condition $\left (r-a\right) \gg k_{\s F}^{-1}$, which we have
already assumed to be met. Still, this value is much bigger than one,
which allows us to replace the summation over $m$ by the integration.
The final result for the oscillating part of $n(r)$ reads

\begin{equation}
\label{deltaN}
\frac{\delta n(r)}{n_0}=-\left(\frac{32ar}{\pi}\right)^{1/2}\!\!\!\left(\frac{1}{r+a}\right)
\frac{\sin\Bigl[2k_{\s F}(r-a)-\frac{\pi}{4}    \Bigr]}{\Bigl[2k_{\s F}(r-a)\Bigr]^{3/2}}.
\end{equation}
In Eq.~(\ref{deltaN}) we have restored the numerical factor. In the domain
$k_{\s F}^{-1} \ll (r-a) \ll a$, Eq.~(\ref{deltaN}) reproduces the Friedel oscillations
Eq.~(\ref{Wall})
 from a hard wall, while for $(r-a)\gg a$ the oscillations
fall off as $1/r^2$ like for a point-like impurity.

\section{Scattering in the absence of interactions}

\subsection{Basic relations}
The general expression for the scattering cross section reads
\begin{equation}
\label{general}
\sigma=\int\limits_0^{2\pi}d\varphi \vert f(\varphi)\vert^2,
\end{equation}
where $f(\varphi)$ is the scattering amplitude. It is
related to the scattering phases, $\delta_m^{(0)}$, as follows\cite{Landau}

\begin{equation}
\label{amplitude}
f(\varphi)=
\frac{1}{\left(2\pi i k \right)^{1/2}}
\sum_{m=-\infty}^{\infty}
\left(e^{2i\delta_m^{(0)}}-1\right)e^{im\varphi}.
\end{equation}
Scattering phases are determined by the long-distance behavior of
the radial wave functions
\begin{equation}
\label{larger}
R_{m,k}(r)\Big\vert_{r\rightarrow \infty}\propto \frac{1}{r^{1/2}}\cos\left(kr-\frac{m\pi}{2}-\frac{\pi}{4}+\delta_m^{(0)}   \right),
\end{equation}
where $k=\frac{1}{\hbar}\left(2mE\right)^{1/2}$ is the wave vector.
Performing the angular integration, one recovers the textbook result\cite{Landau}
\begin{equation}
\label{full}
\sigma=\frac{4}{k}\sum_{m=-\infty}^{\infty}\sin^2\delta_m^{(0)}.
\end{equation}
The quantity that enters the conductivity is not $\sigma$
but the transport cross section defined as
\begin{equation}
\label{general1}
\sigma_{\text{tr}}=\int\limits_0^{2\pi}d\varphi (1-\cos\varphi) \vert f(\varphi)\vert^2.
\end{equation}
Substituting Eq.~(\ref{amplitude}) into Eq.~(\ref{general1})
and integrating over $\varphi$, one obtains
\begin{equation}
\label{integrated}
\sigma_{\text{tr}}=\frac{2}{k}\sum_{m=-\infty}^{\infty}
\sin^2\left(\delta_m^{(0)}-\delta_{m+1}^{(0)}\right).
\end{equation}

\section{Hard disk}

\subsection{Classical calculation}

\begin{figure}
  \includegraphics[scale=0.35]{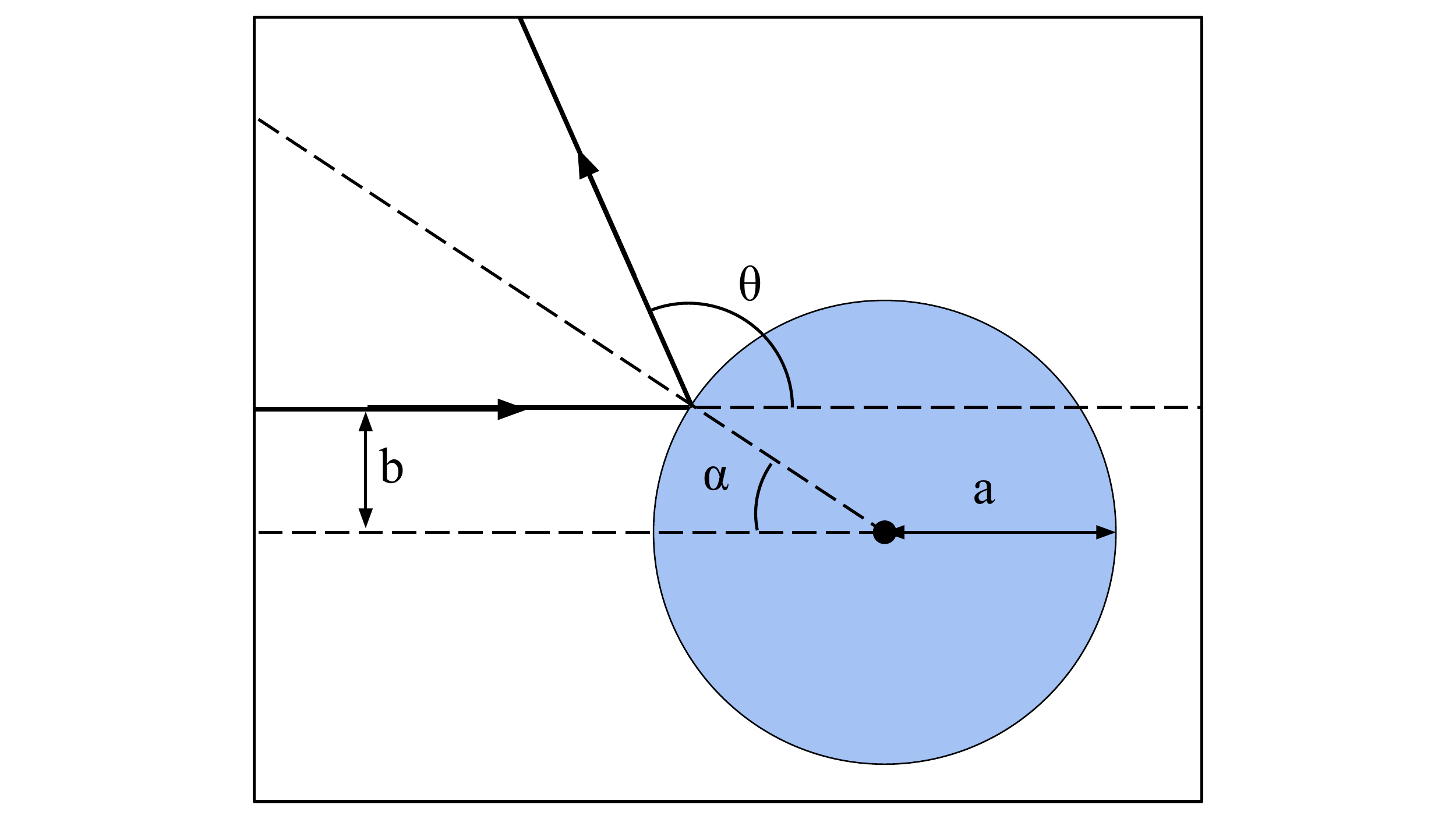}
  \caption{\label{fig:cartoon}(Color online) A schematic of classical hard-disk scattering. A particle with impact parameter $b$ incident on a target with radius $a$. The scattering angle $\theta$ is related to the impact parameter $b$ through Eq.~(\ref{classical1}).}
\end{figure}
Within a classical picture, a particle  incident on the disk with impact parameter, $b<a$, is reflected from the boundary, see Fig. {\ref{fig:cartoon}}. The scattering angle, $\theta$, is related to $b$ as
\begin{equation}
\label{classical1}
b(\theta)=-a\cos\left(\frac{\theta}{2}\right).
\end{equation}
Substituting Eq.~(\ref{classical1}) into the definitions of the full and transport cross sections
\begin{equation}
\label{classical2}
\sigma =\int\limits_0^{2\pi}d\theta \Big\vert\frac{\partial b}{\partial \theta}\Big\vert,~~~~
\sigma_{\text{tr}} =\int\limits_0^{2\pi}d\theta \left(1-\cos\theta\right) \Big\vert\frac{\partial b}{\partial \theta}\Big\vert,
\end{equation}
we get
\begin{align}
\label{classical3}
\sigma=\frac{a}{2}\int\limits_0^{2\pi} d\theta\sin\left(\frac{\theta}{2}\right)=2a,
\nonumber\\
\sigma_{\text{tr}}=\frac{a}{2}\int\limits_0^{2\pi} d\theta\left(1-\cos\theta \right)\sin\left(\frac{\theta}{2}\right)=\frac{8}{3}a.
\end{align}
The above calculation suggests that the classical cross section is equal to the diameter of the disk. Although this diameter is much bigger than the de Broglie wave length, the result obtained neglecting the diffraction effects is   not supported
by the quantum calculation.
\subsection{Quantum calculation}

For a hard disk of a radius, $a$, the  form of
the radial wave function at $r>a$ is the  linear combination
\begin{equation}
\label{phases}
R_{m,k}(r)=\cos\delta_m J_m(kr)+\sin \delta_m N_m(kr),
\end{equation}
of the Bessel and the Neumann functions, $J_m(z)$ and $N_m(z)$, which are the free solutions of the Schr{\"o}dinger equation.
Then the exact expression for the phases, $\delta_m$, which follows from the condition $R_{m,k}(a)=0$, reads
\begin{equation}
\label{sin2phases}
\sin^2\delta_m^{(0)}=\frac{J_m^2(ka)}{J_m^2(ka)+N_m^2(ka)}.
\end{equation}

\begin{figure}
  \includegraphics[scale=0.25]{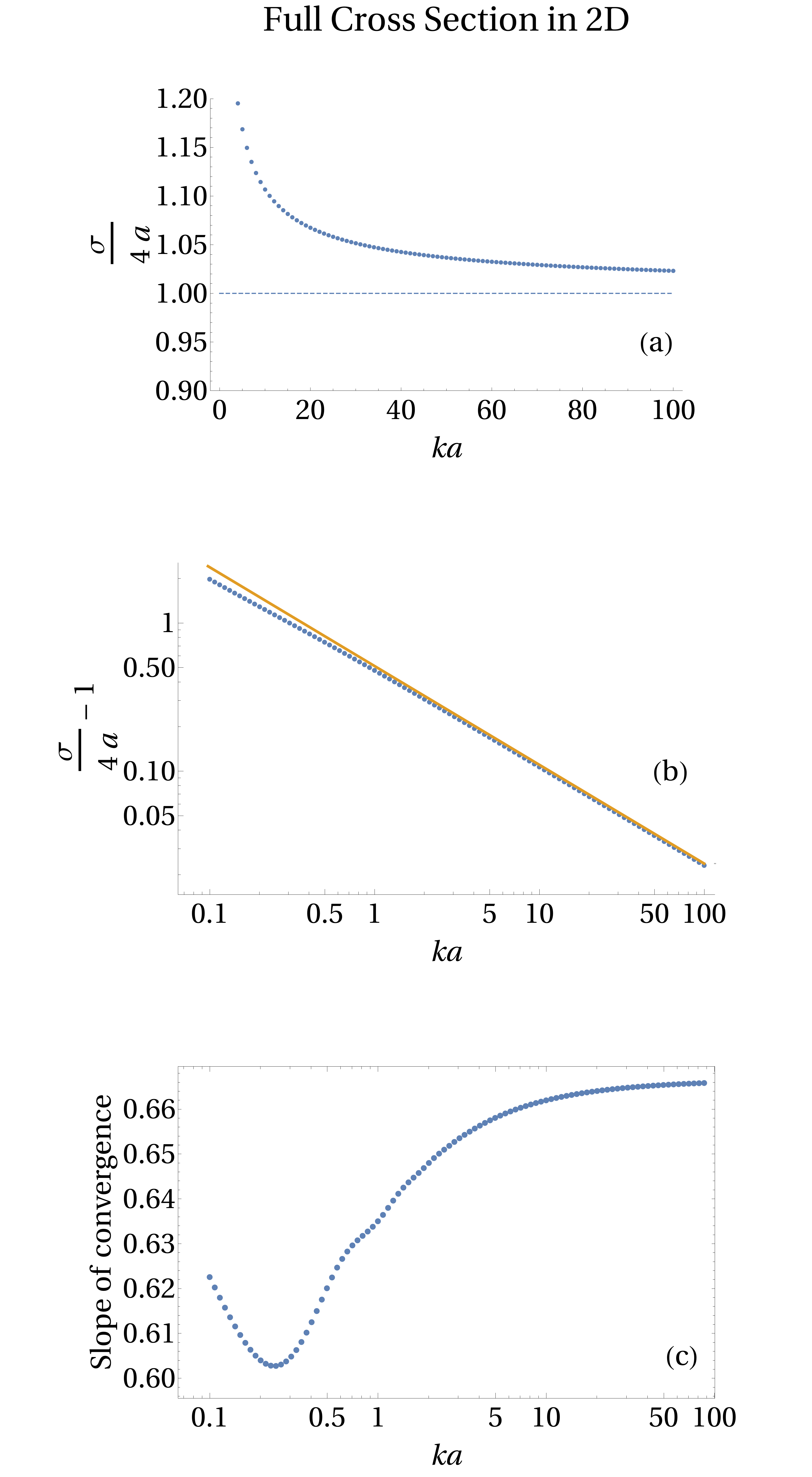}
  \caption{\label{fig:Full2D} (Color online) Full scattering cross section  is calculated numerically from Eqs.~(\ref{full}) and (\ref{sin2phases}). In (a), the approach to the asymptotic value  $\sigma=4a$ is shown. In (b), the log-log plot of $\sigma(ka)$ calculated numerically (blue curve) is compared to the theoretical prediction Eq.~(\ref{correction}) (yellow curve).
  In (c), the slope $\frac{d \sigma}{d \ln ka}$ is plotted versus $ka$.  In agreement with theory, Eq.~(\ref{correction}) the slope  approaches $2/3$ at large $ka$. The minimum at $ka\approx 0.2$ originates from $m=0$ term, Eq.~(\ref{smallks}).}
\end{figure}

At small $ka\ll 1$ the sum Eq.~(\ref{full}) is dominated by the
first term for which $J_0(ka)\approx 1$ and $N_0(ka)\approx \frac{2}{\pi}\ln(ka)$.
Then Eq.~(\ref{full}) takes the form
\begin{equation}
\label{smallks}
\frac{\sigma}{4a}\Big\vert_{ka \ll 1}\approx \frac{\pi^2}{4ka\ln^2(ka)}.
\end{equation}
The right-hand side has a minimum at $ka\approx 0.25$. At this $ka$ the terms with higher $m$ in Eq.~(\ref{full}) are important, leading to the
decay of $\sigma$ with energy. Still, the minimum in $m=0$ term manifests
itself as a shallow minimum in the derivative $\frac{d\sigma}{d (ka)}$, as illustrated by the numerical calculation, see Fig.~\ref{fig:Full2D}.

Upon increasing energy the fall off of the cross section saturates at
$\frac{\sigma}{4a}\approx 1$, which corresponds to replacement of $\sin^2\delta_m$
by $1/2$ for $m<ka$ and by zero for $m>ka$. This value exceeds twice the classical result Eq.~(\ref{classical3}),
which is a well-known effect in 3D, see e.g. Ref. \onlinecite{Landau}.

Next we will study the quantum correction to the scattering cross section, which determines the law of approach of $\sigma$ to the saturation value in the limit $ka \gg 1$. The main point is that this approach is dominated by a narrow domain of
momenta $\left(ka-m\right)\ll ka$.
To capture the quantum correction analytically,  we infer the expression for the phases by comparing
the semiclassical form Eq.~(\ref{semiclassical})
of the radial wave functions to the asymptote Eq.~(\ref{larger}).
The integral in the argument of sine can be evaluated analytically
\begin{align}
I_m \!&= \int\limits_a^r \left(k^2 - \frac{m^2}{{r'}^2}\right)^{1/2}\! dr' \nonumber \\
&= \left(k^2r^2 - m^2\right)^{1/2} - \left(k^2a^2 - m^2\right)^{1/2}\nonumber \\
&- m\arctan\Bigl[\left(\frac{kr}{m}\right)^2 - 1\Bigr]^{1/2}
\!\!\!+ m \arctan\Bigl[\left(\frac{ka}{m}\right)^2 - 1\Bigr]^{1/2}\!\!\!.
\end{align}
Taking the limit $r \rightarrow \infty$ we obtain

\begin{align}
  \label{integralresult}
  I_m &\approx kr - \frac{m \pi}{2} - \left[k^2a^2 - m^2\right]^{1/2}\nonumber \\
  &+ m \arctan\left[\left(\frac{ka}{m}\right)^2 - 1\right]^{1/2}.
\end{align}
With this $I_m$, the semiclassical expression Eq.~(\ref{semiclassical}) matches the asymptote
Eq.~(\ref{larger}) for the following choice of the scattering phases

\begin{align}
\label{bigmphases}
&\frac{3\pi}{4}-\delta_m^{(0)} \nonumber\\
&= m\Biggr\{\left[\left(\frac{ka}{m}\right)^2 - 1\right]^{1/2}
\mkern-18mu -  \arctan\left[\left(\frac{ka}{m}\right)^2 - 1\right]^{1/2}\Biggr\}.
\end{align}
In the domain $\left(ka-m \right)\ll ka$ the argument of arctangent is small,
which allows to use the expansion $z - \arctan z = \frac{1}{3}z^3$, after which
Eq.~(\ref{bigmphases}) simplifies to
\begin{equation}
\label{PHASES}
\frac{3\pi}{4}-\delta_m^{(0)} \approx \frac{(2m_1)^{3/2}}{3(ka)^{1/2}},
\end{equation}
where $m_1=ka-m$ characterizes the proximity of angular momentum to $ka$.
From here we get

\begin{equation}
\label{sin}
  \sin^2 \delta_m^{(0)} = \frac{1}{2} + \frac{1}{2}
  \sin\Biggl[\frac{2(2m_1)^{3/2}}{3(ka)^{1/2}}\Biggr].
\end{equation}
The argument of sine defines the characteristic $m_1\sim (ka)^{1/3}$, which is much
smaller than $ka$, as was assumed above. On the other hand, in the domain $ka \gg 1$,
this characteristic value is much bigger than $1$, which allows one to replace the summation
over $m_1$ by integration. This yields

\begin{align}
\label{correction}
\frac{\sigma}{4a}-1&=\int\limits_0^{\infty}\frac{dm_1}{ka}\sin\Biggl[\frac{2(2m_1)^{3/2}}{3(ka)^{1/2}}\Biggr] \nonumber\\
&=\frac{1}{2}\left(\frac{2}{3k^2a^2}   \right)^{1/3}\!\!\int\limits_0^{\infty}\frac{dz\sin z}{z^{1/3}} \nonumber\\
&=\frac{\pi}{\Gamma(\frac{1}{3})} \frac{1}{12^{1/3}} \frac{1}{(ka)^{2/3}} = \frac{\alpha}{(ka)^{2/3}},
\end{align}
where $\alpha = \frac{\pi}{\Gamma(\frac{1}{3})} \frac{1}{12^{1/3}} \approx 0.51$.

The right-hand side of Eq.~(\ref{correction}) is the amount by which the cross section
at finite energy exceeds the limiting value, $\sigma =4a$.

\subsection{Transport cross section in 2D}

The contribution to $\sigma$ and to $\sigma_{\text{tr}}$
from big momenta, $|m|>ka$, is exponentially small.
For $|m|\ll ka$ the asymptotic expression for the phases is\cite{Lapidus}
\begin{equation}
\label{smallmphases}
\delta_m^{(0)}=-\arctan \Biggl(\frac{J_m(ka)}{N_m(ka)}\Biggr)
\approx ka-\frac{\pi}{2}\left(m+1\right)-\frac{\pi}{4}.
\end{equation}
A more general expression for  $\delta_m^{(0)}$ which is valid for {\em all}
$m$ smaller than $ka$ can be inferred
from the semiclassical expression Eq.~(\ref{semiclassical})
by calculating the integral in the argument of sine explicitly.
This yields
\begin{equation}
\label{more}
\delta_m^{(0)}=\left[\left(ka\right)^2-m^2\right]^{1/2}-
m\arctan\left[\left(\frac{ka}{m}\right)^2-1\right]^{1/2}+\frac{3\pi}{4}.
\end{equation}
In calculating the full cross section it is sufficient to replace $\sin^2\delta_m^{(0)}$ by $1/2$ in all $2ka+1$ terms in
Eq.~(\ref{full}) for which the phase is big. Upon doing so, the standard result\cite{Lapidus}
\begin{equation}
\label{FULL}
\sigma \approx 4a,
\end{equation}
is reproduced. As in the 3D case,\cite{Landau}
the result Eq.~(\ref{FULL}) does not contain
the wavelength of the incident electron. Still, the full cross section exceeds
twice the geometric cross section.

\begin{figure}
  \includegraphics[scale=0.25]{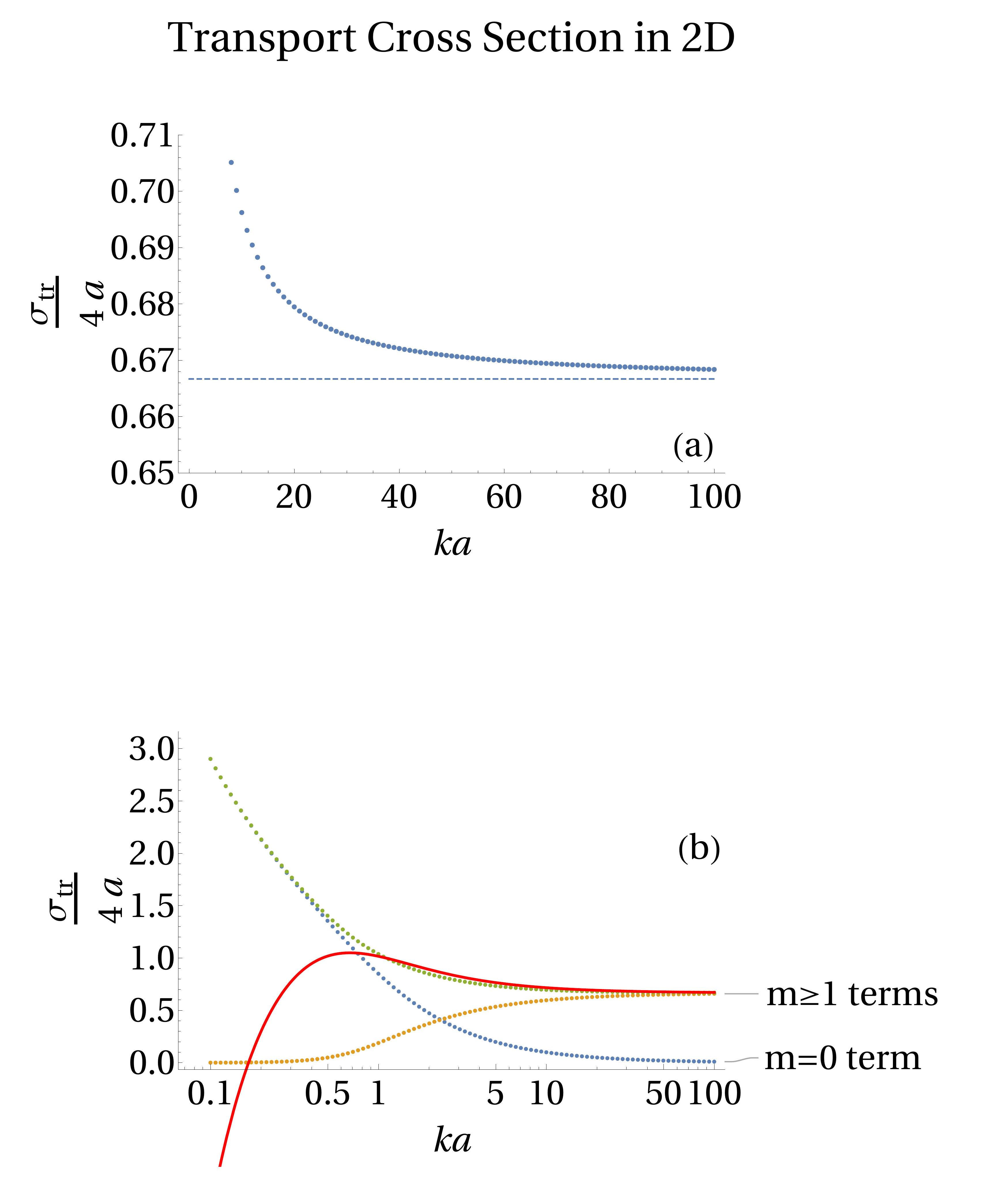}
  \caption{\label{fig:Transport2D} (Color online) The transport cross section in two dimensions is calculated numerically from Eq.~(\ref{T2DSumDenominator}). In (a), the approach to the asymptotic value of $8a/3$ is shown. In (b), a plot of the different contributions to the sum in Eq.~(\ref{T2DSumDenominator}). The blue dots represent the contribution from the terms with $m \geq 1$. The yellow dots represent the $m = 0$ term. The green dots are the sum of the blue and yellow. The red curve is a plot of Eq.~(\ref{T2Dfinal}) and is a good approximation for $ka > 1$.}
\end{figure}

The calculation of the transport cross section is a more delicate task,
since the difference $\delta_m^{(0)}-\delta_{m+1}^{(0)}$ is a slow
function of $m$. Indeed, from Eq.~(\ref{more}) we get
\begin{equation}
\label{correct}
\delta_m^{(0)}-\delta_{m+1}^{(0)} \approx -\frac{\partial \delta_{m}^{(0)}}{\partial m}=
\arctan \left(\frac{k^2a^2}{m^2}-1\right)^{1/2}.
\end{equation}
From the above relation we find
\begin{equation}
\label{correct}
\sin^2\Bigl(\delta_m^{(0)}-\delta_{m+1}^{(0)}\Bigr) \approx 1-\Bigl(\frac{m}{ka}\Bigr)^2.
\end{equation}
Then the summation over $m$ leads to the following result
for the  transport scattering cross section
\begin{equation}
\label{T2D_asymp_value}
\sigma_{\text{tr}}\approx \frac{8}{3}a.
\end{equation}
Unlike the full cross section, this result coincides coincides with
the transport cross section calculated classically.
To study the quantum correction to $\sigma_{\text{tr}}$, we express the difference $\delta_m^{(0)} -\delta_{m+1}^{(0)}$
in terms of the Bessel functions. This yields

\begin{align}
  \mkern-18mu \frac{\sigma_\text{tr}}{4a} &= \frac{1}{2ka} \sum_{m=-\infty}^{\infty} \sin^2\left( \delta_m^{(0)} -  \delta_{m+1}^{(0)}\right)\nonumber\\
  \label{T2DBesselForm}
  &= \frac{1}{2ka}
  \!\!\!\sum_{m=-\infty}^{\infty} \!\!\frac{\left[J_m(ka)N_{m+1}(ka) - N_m(ka)J_{m+1}(ka)\right]^2}{\left[J^2_m(ka) + N^2_m(ka)\right]\left[J_{m+1}^2(ka)+N_{m+1}^2(ka)\right]}.
\end{align}
The numerator of Eq.~(\ref{T2DBesselForm}) can be
greatly simplified upon using the relation\cite{{Bateman}}
\begin{equation}
  \label{numeratorConstant}
  J_m(ka)N_{m+1}(ka) - N_m(ka)J_{m+1}(ka) = -\frac{2}{\pi k a}.
\end{equation}
With this simplification Eq.~(\ref{T2DBesselForm}) assumes the form

\begin{multline}
  \label{T2DSumDenominator}
  \frac{\sigma_{\text{tr}}}{4a} = \frac{4}{\pi^2(ka)^3}\\
\times  \sum_{m=0}^{\infty} \Biggr(\frac{1}{\left[J^2_m(ka) + N^2_m(ka)\right]\left[J_{m+1}^2(ka)+N_{m+1}^2(ka)\right]}\Biggr).
\end{multline}
The brackets in the denominator
can be analyzed using the
integral representation\cite{Bateman}

\begin{align}
\label{representation}
  &J^2_m(ka) + N^2_m(ka) \nonumber \\
  &= \frac{8}{\pi^2} \int\limits_0^\infty \cosh(2mt)
  K_0 \left(2ka \sinh t\right) dt,
\end{align}
where $K_0(z)$ is the Macdonald function.
 Since the product $ka$ is big, the integral is dominated by small $t\ll 1$, which allows to replace $\sinh t$ by
 $t$. To perform the integration over $t$ it is convenient to use the following representation of the Macdonald function
 \begin{equation}
 \label{Macdonald}
 K_0(2kat)=\int\limits_1^{\infty}\frac{ds}{\left(s^2-1  \right)^{1/2}}\exp\bigl(-2kats\bigr).
 \end{equation}
Substituting Eq.~(\ref{Macdonald}) into Eq.~(\ref{representation}) and integrating over $t$, we get

\begin{equation}
\label{simplified}
J^2_m(ka) + N^2_m(ka)=\frac{4}{\pi^2ka}\int\limits_1^{\infty}
\frac{ds s}{\left(s^2-1\right)^{1/2}
\left[s^2-\left(\frac{m}{ka}    \right)^2    \right]}.
\end{equation}
Now the evaluation of the integral is elementary and is achieved by the substitution $s=(1+u^2)^{1/2}$. The result reads
\begin{equation}
\label{simplified1}
J^2_m(ka) + N^2_m(ka)=\frac{2}{\pi ka\left(1-\frac{m^2}{k^2a^2}   \right)^{1/2}}.
\end{equation}
This result applies for $m<ka$. For $m>ka$ the integral
Eq.~(\ref{simplified}) is zero. Using the relation Eq.~(\ref{simplified1}), the expression Eq.~(\ref{T2DSumDenominator})
for the transport cross section can be cast in the form

\begin{equation}
  \label{T2D_wSum}
  \frac{\sigma_\text{tr}}{4a} = \frac{1}{ka}\sum_{m=0}^\infty \left(1 - \frac{m^2}{k^2a^2} \right)^{1/2}\left(1 - \frac{(m+1)^2}{k^2a^2} \right)^{1/2}.
\end{equation}
If we neglect the difference between $(m+1)$ and $m$ in the second bracket, the product of the brackets will reduce to $\left( 1 - \frac{m^2}{k^2a^2}\right)$. Then the summation over $m$
will reproduce the result Eq.~(\ref{T2D_asymp_value}). Thus, the quantum correction to the transport cross section is due to the difference between the first and second brackets. To account for this difference
we expand the second bracket as follows

\begin{align}
&\left(1-\frac{\left(m+1\right)^2}{k^2a^2}\right)^{1/2}\nonumber
 \\
 \label{expansion_of_mplus1}
&\approx\left(1 - \frac{m^2}{k^2a^2} \right)^{1/2}-\frac{m}{k^2a^2\left(1 - \frac{m^2}{k^2a^2} \right)^{1/2}}.
\end{align}
Substituting this expansion into Eq.~(\ref{T2D_wSum}) and performing
summation over $m$, we arrive to the corrected expression for $\sigma_{\text{tr}}$
\begin{equation}
\label{sizeeffect}
  \frac{\sigma_\text{tr}}{4a} = \frac{2}{3} - \frac{1}{2ka}.
\end{equation}
We see that the correction is negative suggesting that the approach
of $\sigma_\text{tr}$ to the limiting value is ``from below". This is the
result of the expansion Eq.~(\ref{expansion_of_mplus1}) underestimating the $m=0$
term. Incorporating this term explicitly, we obtain

\begin{multline}
  \label{T2Dfinal}
  \frac{\sigma_\text{tr}}{4a} = \frac{2}{3} - \frac{1}{2ka}\\
   +\frac{4}{\pi^2(ka)^3\left[J^2_0(ka) + N^2_0(ka)\right]\left[J_{1}^2(ka)+N_{1}^2(ka)\right]}.
\end{multline}
As seen in Fig.~\ref{fig:Transport2D},
Eq.~(\ref{T2Dfinal})
leads to the approach of $\sigma_\text{tr}$ to the limiting value ``from above" for $ka > 1$. Moreover, it reproduces correctly the result of numerical calculation.

\section{Incorporating interactions}

We will follow a transparent procedure of incorporating the interactions
which was put forward in Ref. \onlinecite{Rudin1997} for the interaction correction
to the density of states and then adopted in Refs.
\onlinecite{Zala}, \onlinecite{Gornyi2006} for the interaction correction to the conductivity.
The main message of Ref. \onlinecite{Rudin1997}, see also
Ref. \onlinecite{Yue}, is that Friedel oscillations of the electron density
translate into the oscillating Hartree potential

\begin{equation}
\label{Hartree}
V_H({\bf r})=\int d{\bf r'} V\left({\bf r}-{\bf r'}\right)\delta n({\bf r'}).
\end{equation}
Here $V\left({\bf r}-{\bf r'}\right)$ is the screened electron-electron
interaction potential. Additional scattering of the incident electron from
the potential $V_H({\bf r})$ modifies the scattering phases $\delta_m^{(m)}$.
In order to calculate the corresponding corrections to the phases, we employ the
procedure described
e.g. in Ref.~\onlinecite{Landau}.

As a first step, instead of the wave function $R_{m,k}^{(0)}(r)$, we
introduce an auxiliary function  $\chi_{m,k}^{(0)}(r)$ defined as
\begin{equation}
\label{chi}
R_{m,k}^{(0)}(r)=\frac{\chi_{m,k}^{(0)}(r)}{\left( kr \right)^{1/2}}.
\end{equation}
The function $\chi_{m,k}^{(0)}$ satisfies the equation
\begin{equation}
\label{radial1}
\frac{d^2}{dr^2}\chi_{m,k}^{(0)} +\left[k^2-\frac{\left(m^2-\frac{1}{4}\right)}{r^2}    \right]\chi_{m,k}^{(0)}=0.
\end{equation}
In the presence of interactions, the potential $V_H(r)$ adds to the centrifugal
potential, so that Eq.~(\ref{radial1}) assumes the form
\begin{equation}
\label{radial2}
\frac{d^2}{dr^2}\chi_{m,k} +\left[k^2+V_{H}(r)-\frac{\left(m^2-\frac{1}{4}\right)}{r^2}    \right]\chi_{m,k}=0.
\end{equation}
Multiplying Eq.~(\ref{radial1}) by $\chi_{m,k}(r)$ and Eq.~(\ref{radial2})
by $\chi_{m,k}^{(0)}(r)$ and subtracting, we arrive to the relation
\begin{equation}
\label{difference}
\frac{d}{dr}\Bigg[ \frac{d\chi_{m,k}}{dr}\chi_{m,k}^{(0)}- \frac{d\chi_{m,k}^{(0)}}{dr}\chi_{m,k} \Bigg]=-V_{H}\chi_{m,k}^{(0)}\chi_{m,k}.
\end{equation}
This relation allows to find the interaction-induced correction to the scattering phase. Since this correction is small, the product
$\chi_{m,k}^{(0)}\chi_{m,k}$ in the right-hand side can be replaced
by $\left(\chi_{m,k}^{(0)}\right)^2$. Then the integration of Eq.~(\ref{difference})
from $a$ to $\infty$ yields

\begin{equation}
\label{Integrated}
\Delta \delta_m=-\int\limits_a^{\infty} dr~ V_{H}(r) \left(\chi_{m,k}^{(0)}\right)^2.
\end{equation}
The correction $\Delta \delta_m$ to the scattering phases
give rise to the following interaction corrections to
the full and to the transport scattering cross sections
\begin{equation}
\label{Correction0}
\delta \sigma=\frac{4}{k}\sum_{m}\sin\left(2\delta_m^{(0)}\right)
\Delta\delta_m,
\end{equation}
\begin{equation}
\label{Correction1}
\delta \sigma_{\text{tr}}=\frac{2}{k}\sum_{m}\sin \Biggl[ 2\left(\delta_m^{(0)}-\delta_{m+1}^{(0)}   \right)\Biggr] \Bigl(\Delta\delta_m-\Delta\delta_{m+1}  \Bigr).
\end{equation}

Next we express the factors $\sin 2\delta_m^{(0)}$ and
$\sin 2\left(\delta_m^{(0)}-\delta_{m+1}^{(0)}   \right)$
in terms of the Bessel functions
\begin{equation}
\label{identity0}
\sin 2\delta_m^{(0)}=\frac{2J_m(ka)N_m(ka)}{ J_m^2(ka)+N_m^2(ka)},
\end{equation}


\begin{align}
\label{identity1}
&\sin 2\left(\delta_m^{(0)}-\delta_{m+1}^{(0)}\right)
\nonumber\\
&=\frac{J_m(ka)N_{m+1}(ka)- N_m(ka)J_{m+1}(ka)}{ J_m^2(ka)+N_m^2(ka)   }
\nonumber\\ &\times
\frac{N_m(ka)N_{m+1}(ka)+ J_m(ka)J_{m+1}(ka)}
{ J_{m+1}^2(ka)+ N_{m+1}^2(ka)}.
\end{align}
Using the fact that $ka\gg 1$ we can simplify the expressions for $\delta \sigma$ and $\delta\sigma_{\text{tr}}$ as follows
\begin{equation}
\label{CORRECTION1}
\delta \sigma=-\frac{8\cos(2ka)}{k}\sum_{m=0}^{ka}\left(-1\right)^m\Delta\delta_m,
\end{equation}
\begin{equation}
\label{CORRECTION2}
\delta \sigma_{\text{tr}}=\frac{8}{k}\sum_{m=0}^{ka}\frac{m}{ka}
\Biggl[1-\left(\frac{m}{ka}\right)^2\Biggr]^{1/2}
\left(\Delta\delta_m - \Delta\delta_{m+1}\right).
\end{equation}
We see that the two expressions are very different. Since $\Delta\delta_m$ is a smooth function
of $m$, the terms in Eq.~(\ref{CORRECTION1}) cancel out. The same smoothness of $\Delta\delta_m$
allows to replace the sum over $m$ in Eq.~(\ref{CORRECTION2}) by the integral
\begin{equation}
\label{continuous}
\delta \sigma_{\text{tr}}=-\frac{8}{k}\int\limits_0^{ka}dm\left(\frac{m}{ka}\right)
\Biggl[1-\left(\frac{m}{ka}\right)^2\Biggr]^{1/2}\frac{\partial\Delta\delta_m}{\partial m}.
\end{equation}
Next we argue that relevant values of $m$ are much smaller than $ka$. This allows to replace
the square bracket by $1$ and transform Eq.  (\ref{continuous}) by parts. This yields
\begin{equation}
\label{byparts}
\delta \sigma_{\text{tr}}=\frac{8}{k^2a}\int\limits_0^{\infty}dm \left(\Delta\delta_m\right).
\end{equation}

To analyze the dependence of $\Delta\delta_m$ given by
Eq.~(\ref{Integrated}) on the wave vector, $k$, of the incident electron
we recall that the potential, $V_H(r)$, is proportional to electron density
given by Eq.~(\ref{density}). To pinpoint the origin of the anomaly at $k=k_{\s F}$
it is more convenient to study the derivative $\frac{\partial \Delta\delta_m}{\partial k_{\s F}}$.
Within a factor, this derivative is given by
\begin{equation}
\label{derivative}
\frac{\partial \Delta\delta_m}{\partial k_{\s F}}\propto \int\limits_a^{\infty}
dr \Bigl[rR_{m,k}^2(r)\Bigr]\sum_{m'}R_{m',k_{\s F}}^2(r).
\end{equation}
From the semiclassical form of the radial wave function Eq.~(\ref{semiclassical})
we conclude that the product $R_{m,k}^2(r)R_{m',k_{\s F}}^2(r)$ contains a slow part
\begin{equation}
\label{slowpart}
\frac{\cos2\Bigl\{\int\limits_a^rdr'\Bigl[\left(k^2-\frac{m^2}{r^2}\right)^{1/2}
 -\left(k_{\s F}^2-\frac{m'^2}{r^2}\right)^{1/2}\Bigr]\Bigr\}}
 {\left(k^2r^2-\frac{m^2}{r^2}   \right)^{1/2}\left( k_{\s F}^2r^2-m'^2\right)^{1/2}}.
\end{equation}
Since we assumed that $m$ and $m'$ are both much smaller than $ka$, the above expression
can be simplified as follows
\begin{equation}
\label{slowpart1}
\frac{\cos2\Bigl[(k-k_{\s F})(r-a)-\frac{m^2-m'^2}{2k}\Bigl(\frac{1}{a}-\frac{1}{r}\Bigr)\Bigr]}
 {k_{\s F}^2r^2}.
\end{equation}
It is natural to measure the radial coordinate, $r$, from $r=a$. Combining Eqs.
(\ref{byparts}), (\ref{derivative}),
 and (\ref{slowpart1}), we arrive to the following expression for
the derivative of $\delta\sigma_{\text{tr}}$ with respect to $k_{\s F}$
\begin{equation}
\label{derivative_simplified}
\frac{\partial \delta\sigma_{\text{tr}}}{\partial k_{\s F}}\propto \ba
\int\limits_0^{\infty}\ba dm\ba \int\limits_0^{\infty}\ba dm'\ba\int\limits_0^{\infty}\frac{d\rho}{\rho+a}
\cos 2\Bigl[(k-k_{\s F})-\frac{m^2-m'^2}{2ka(\rho+a)}\Bigr]\rho.
 \end{equation}
We can now perform the integration over $m$ and $m'$ explicitly. This yields
\begin{equation}
\label{overm}
\frac{\partial \delta\sigma_{\text{tr}}}{\partial k_{\s F}}\propto
2\pi ka\int\limits_0^{\infty}\frac{d\rho}{\rho}
\cos2\Big[\left(k-k_{\s F}\right)\rho\Big].
\end{equation}
Since the expression for the Friedel oscillations applies for $\rho=(r-a)\gg k_{\s F}^{-1}$,
the lower limit in the integral should be chosen to be $\rho \sim k_{\s F}^{-1}$. Thus, we arrive
to the final result
\begin{equation}
\label{final}
\frac{\partial \delta\sigma_{\text{tr}}}{\partial k_{\s F}}\propto \ln\Bigl(\frac{k_{\s F}}{k-k_{\s F}}   \Bigr)
=\ln\Bigl(\frac{E_{\s F}}{\varepsilon}   \Bigr).
\end{equation}
From Eq.~(\ref{final}) we conclude that the interaction correction to the transport cross section
has the form $\delta\sigma_{\text{tr}} \propto \varepsilon \ln\Bigl(\frac{E_{\s F}}{\varepsilon}   \Bigr)$.
Recall, that for point-like impurity\cite{Zala} the interaction correction has the form
$\delta\sigma_{\text{tr}} \propto \varepsilon$. Thus, the enhancement of the interaction correction
in the case of scattering from the disk amounts to the logarithmic factor.
\begin{figure}
  \includegraphics[scale=0.25]{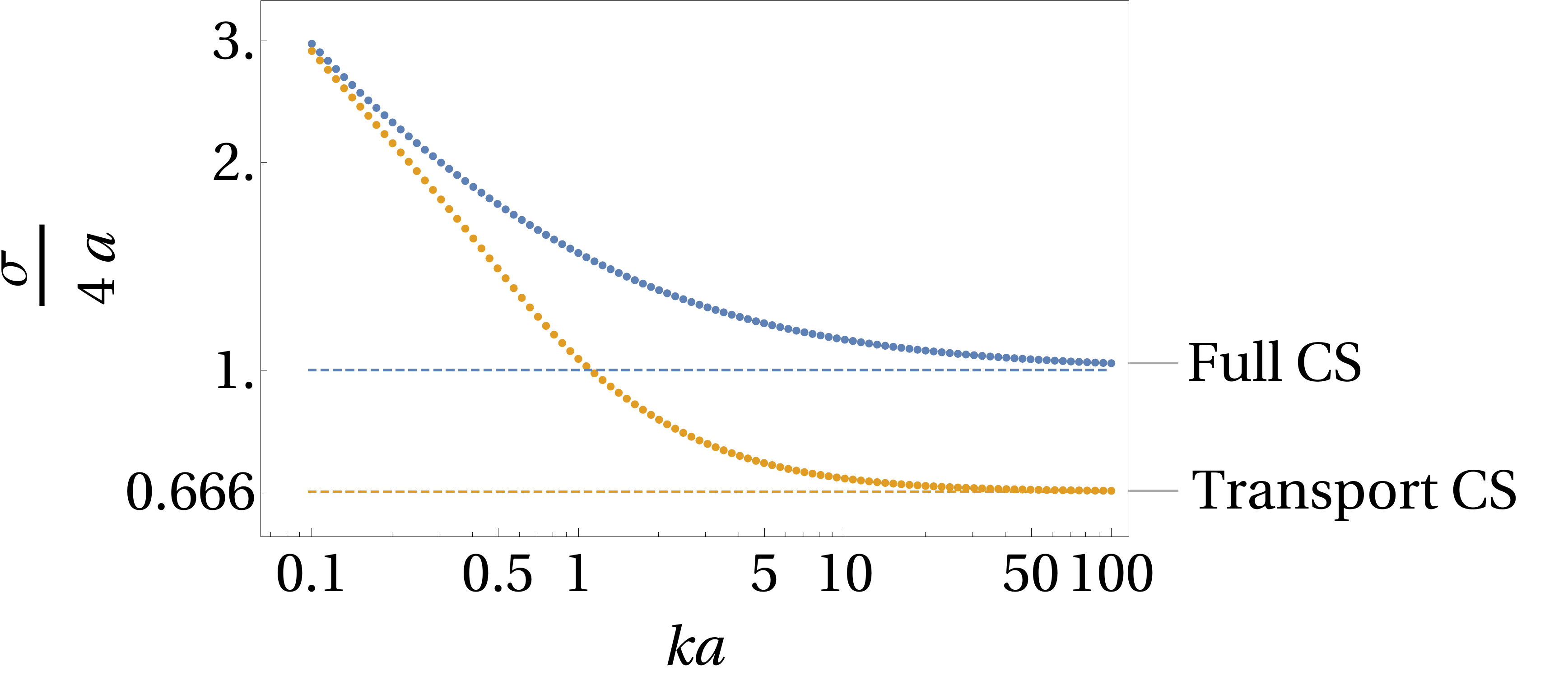}
  \caption{\label{fig:comp2Dsigma} (Color online) Full cross section (blue curve) and scattering cross section (yellow curve)
  are plotted in log-log scale from Eqs.~(\ref{full}) and (\ref{integrated}), respectively.
  With increasing the dimensionless parameter, $ka$, the transport cross section
  approaches the limiting value, $\frac{8a}{3}$, faster than the full cross section approaches the limiting
  value, $4a$.}
\end{figure}

\section{Concluding remarks}

\noindent ({\em i}) In Fig.~\ref{fig:comp2Dsigma} the numerical results for the full and transport
scattering cross sections are presented in log-log scale. It is apparent that
$\sigma$ and  $\sigma_{\text{tr}}$ approach their limiting values $\sigma=4a$ and
$\sigma_{\text{tr}}=\frac{8}{3}a$, respectively, at very different rates.
Naturally, at small $ka\ll 1$ both cross sections coincide. However, at large $ka$,
the transport cross section saturates much faster than the full cross section.
On the contrary, we have demonstrated that while the interaction correction to $\sigma_{\text{tr}}$ is
a singular function of $k-k_{\s F}$, the interaction correction to $\sigma$ vanishes.

\noindent ({\em ii}) Integration over momenta $m$ and $m'$ in Eq.~(\ref{derivative_simplified})
leads to the slow energy dependence of the correction
$\frac{\partial \delta\sigma_{\text{tr}}}{\partial k_{\s F}}$.
Note, that this integration misses a contribution from the terms $m=m'$, which is
the consequence of the discreteness of the angular momentum.
This contribution is comparable to the main contribution Eq.~(\ref{final}) for
the following reason. In integration, the characteristic values of $m$ and $m'$
are $\sim (ka)^{1/2}$, while the number of terms with $m=m'$ is $\sim ka$.
On the other hand, the contribution of these terms possesses a sharp energy dependence.
This contribution can be cast in the form
\begin{equation}
\label{sharp}
\frac{\partial \delta{\tilde\sigma}_{\text{tr}}}{\partial k_{\s F}}
\propto \int\limits_0^{\infty}\frac{dz\cos z}{z+2\left(k-k_{\s F}\right)a}=
\int\limits_0^{\infty}\frac{dz\sin z}{\Bigl[z+2\left(k-k_{\s F}\right)a\Bigr]^2}.
\end{equation}
Eq.~(\ref{sharp}) yields a narrow peak of a width $\left(k-k_{\s F}\right)\sim a^{-1}$ and the
corresponding energy scale $\varepsilon \sim \frac{E_{\s F}}{ka}$. At large $\left(k-k_{\s F}\right)a \gg 1$
the correction Eq.~(\ref{sharp}) falls off as $\frac{1}{4\left(k-k_{\s F}\right)^2a^2}$.
Note also, that the correction Eq.~(\ref{sharp}) originates from the specific behavior of
the Friedel oscillations Eq.~(\ref{deltaN}) from the disk as compared to the Friedel oscillations
from the point impurity. It thus contains the disk radius, $a$. Upon integrating Eq.~(\ref{sharp}),
we get the following shape of the peak in the transport cross section
\begin{equation}
\label{sharp1}
\delta{\tilde\sigma}_{\text{tr}}
\propto
\int\limits_0^{\infty}\frac{dz\sin z}{z+2\left(k-k_{\s F}\right)a}.
\end{equation}
The missing prefactor in Eq.~(\ref{sharp1}) contains the product of the  electron-electron interaction
constant\cite{Zala} and the density of states, $\nu_0$. The first factor originates from the
proportionality between $V_H(r)$ and $\delta n(r)$ in Eq.~(\ref{Integrated}),
while the second factor comes from the sum in Eq.~(\ref{derivative}), which
emerges upon taking the derivative with respect to $k_{\s F}$.

\noindent ({\em iii}) We emphasize the difference between the interaction corrections for the
cases of a point-like impurity and of a big disk with $ka\gg 1$. For a point-like
impurity,\cite{Zala} the corrections to the full and to the transport scattering cross sections
are related as $\delta\sigma_{\text{tr}} =2\delta\sigma$. This can also be seen from Eq.~(\ref{Integrated}).
For $ka\ll 1$, the correction, $\Delta\delta_0$, is much bigger than $\Delta\delta_m$ for $m\neq 0$.
By contrast, for $ka\gg 1$, the correction, $\delta\sigma$, is small in parameter $\frac{1}{ka}$.

\section{Acknowledgements}

N.F. acknowledges the support of the National Science Foundation (NSF) award No. 1950409. M.E.R. was supported by the Department of Energy, Office of Basic Energy Sciences, Grant No. DE-FG02-06ER46313.

\end{document}